\documentclass{article}
\usepackage{natbib}
\usepackage[utf8]{inputenc}
\usepackage{graphicx}
\usepackage{hyperref}
\usepackage{url}
\usepackage{color}
\usepackage{booktabs}
 \usepackage{adjustbox}
\usepackage{fancyhdr}
\usepackage{makecell}
\usepackage[section]{placeins}

\providecommand{\keywords}[1]
{
  \small	
  \textbf{\textit{Keywords---}} #1
}
\usepackage{geometry}
 \usepackage{subfiles}

\title{Estimating and Projecting Air Passenger Traffic during the COVID-19 Coronavirus Outbreak and its Socio-Economic Impact}
\author{Stefano Maria Iacus \and Fabrizio Natale \and Carlos Satamaria  \and Spyridon Spyratos \and Michele Vespe}
 
 \date{%
   \small{ European Commission, Joint Research Centre}\\[2ex]%
    \today
}
\begin{document}

\maketitle

\begin{abstract}
The main focus of this study is to collect and prepare data on air passengers traffic worldwide with the scope of analyze the impact of travel ban on the aviation sector. Based on historical data from January 2010 till October 2019, a forecasting model is implemented in order to set a reference baseline. Making use of airplane movements extracted from  online flight tracking platforms and on-line booking systems,  this study presents  also a first assessment of recent changes in flight activity around the world as a result of the COVID-19 pandemic.
To study the effects of air travel ban on aviation and in turn its  socio-economic, several  scenarios are constructed based on past pandemic crisis and the observed flight volumes. It turns out that, according to this hypothetical scenarios, in the first Quarter of 2020 the impact of aviation losses could have negatively reduced World GDP by 0.02\% to 0.12\% according to the observed data and, in the worst case scenarios, at the end of 2020 the loss could be as high as 1.41-1.67\% and job losses may reach the value of 25-30 millions. Focusing on EU27, the GDP loss may amount to 1.66-1.98\% by the end of 2020 and the number of job losses from 4.2 to 5 millions in the worst case scenarios. Some countries will be more affected than others in the short run and most European airlines companies will suffer from the travel ban.

\section*{Highlights}
\begin{itemize}
   \item This work estimates different scenarios of air passengers loss due to the Covid-19 pandemia.
    \item This work estimates the impact on aviation contraction on GDP growth and jobs losses worldwide and in EU27.
   \item It is observed that in the first quarter of 2020, in most favorable scenarios about 0.02\% to 0.12\% of the world GDP (0.02\% to 0.13\% for EU27) could have been lost and, in the worst case scenarios, these number raise between 1.41\% and 1.67\% (respectively, 1.66\% and 1.98\% for EU27) for the whole year 2020.
   \item Under the hypothesized scenarios the number of potential jobs lost in the aviation sector (direct and indirect) in the first Quarter of 2020, may vary between 310,000 and 2.21 million in most favorable scenarios (resp. 40,000 to 330,000 for the EU27), and between  25.68 and 30.31 million of units in 2020 (respectively, 4.19 and 5 for the EU27).
        \item The results show that air traffic follows dynamics that appear to be geographically correlated to the spreading of the COVID-19 outbreak to different parts of the world. 
    \item According to our estimates, during the week 19-25 March the traffic had dropped globally by 52\% compared to the traffic during the week 31 January-5 February. EU27 was the region worst affected by the decline in activity (down 65\%).    
\end{itemize}
\end{abstract}

\keywords{COVID-19, coronavirus, air passengers data}

\pagebreak

\tableofcontents

\pagebreak
\section{Introduction}
The recent COVID-19 Coronavirus outbreak and the relevant precautionary measures to limit its spreading are having clear impacts on human mobility at global scale. This provoked a reduction of domestic and international volumes of air passenger traffic to and from China in February \citep{iacus2020flight}. Such effects are currently being observed in several regions worldwide. This has clear implications for the aviation industry as well as indirect consequences to several sectors (e.g. tourism) and the economy at large as well as the society. Through the use of historical air traffic data, real time flights tracks and on-line booking systems, this works aims to provide air traffic volume projections following a set of scenarios base on observed and previous crises. These range from a rapid and full recovery to less optimistic scenarios of slower or even incomplete recovery which will depend on the duration and intensity of the lock-down.

The trend of mobility at global scale has been rising over the last decade at a pace that is faster than the global world population growth \citep{recchi2019estimating}. Nevertheless, air traffic flows have been shaped at national and regional scale by shocks due to political instability,  terrorism and economic crises \citep{gabrielli2019dissecting}. In addition, the air traffic industry has shown strong dependency on pandemia outbreaks in the past such as SARS in 2003 and MERS in 2015 \citep{IATApandemic}, with effects that had repercussions at regional and global scale.

The paper is organized as follows: in Section~\ref{sec:data} we present the different data sources. Section~\ref{sec:model} briefly presents the forecasting model used as a baseline reference to estimate the impact of travel ban. Section~\ref{sec:scenarios} presents the different scenarios imagined for this COVID-19 pandemia based on past similar events. Section~\ref{sec:impact} is an attempt to provide a rough estimate of the impact of travel ban wordlwide on aviation directly and on  GDP and job losses consequently. Section~\ref{sec:europe} focuses on the EU27. Finally, Section~\ref{sec:tracking} drills down to the airport level to investigate the actual lock-down measures worldwide.

\section{Data}\label{sec:data}
In this section we present the different data sources used in this work to show the different types of impact on aviation, economy and society.
\subsection{Historical Air Traffic Passenger Data}

\begin{figure}[!htb]
    \centering\includegraphics[width=1\textwidth]{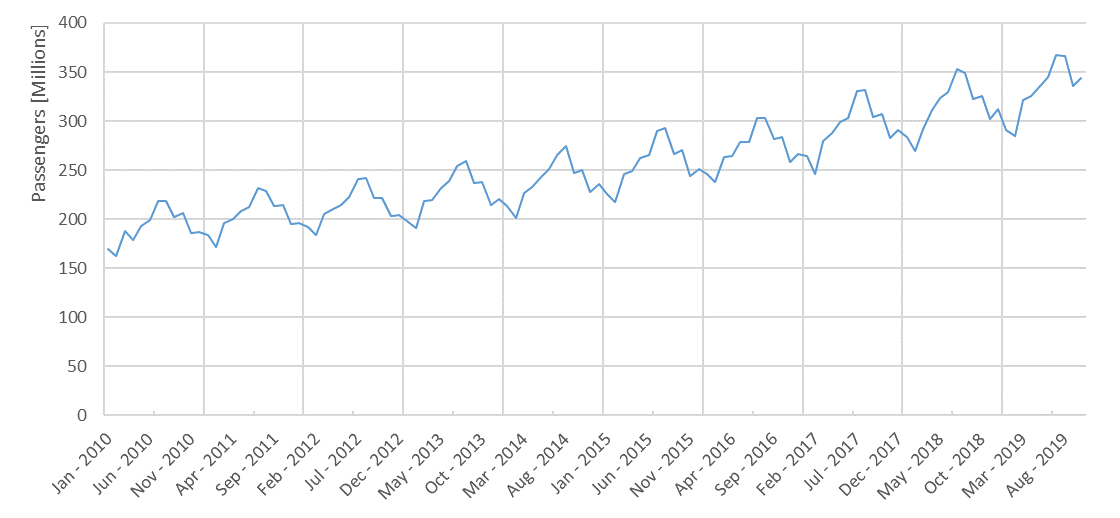}
    \caption{\textit{Aggregated volume of global air traffic passengers from January 2010 to October 2019.}}
    \label{figTREND}
\end{figure}

Historical air passenger data used in this work contain consolidated monthly air traffic information on the number of passengers and the average price per ticket, for every couple of airports around the world for which a direct or indirect connection exists, i.e., the data register the actual flight volumes from each true origin and true final destination, taking into account intermediate stops.  The dataset is processed and provided by SABRE \citep{sabre}. The latest available data point in time at the time of this writing is October 2019, a few months before the outbreak of COVID-19 Coronavirus, and starts on September 2010. The volume of air traffic passengers at global scale over the last decade is shown in Figure \ref{figTREND} where, besides the seasonal effect, an overall increasing trend can be observed. 
As mentioned, the data considered in our analysis consist of routes (international and domestic) from an origin airport to a destination airport around the world, i.e., itinerary which possibly consist of one or more stops. This aspect of the data has advantages and disadvantages. For example, in the previous analysis on China aviation \citep{iacus2020flight}, while all direct flights from China to Italy were banned, using the on-line booking platforms explained later on, it was clearly seen from the data that 1-stop and 2-stop flights could guarantee the connections from China to Italy during February 2020. This aspect is important in the analysis of the pandemic outbreak. The disadvantage is that it is not possible to understand from the data if a route is a direct flight or not, without prior information and it is not possible to disentangle either whether a route contains a mix of direct and non-direct flights.

The routes data set consist of 1,056,097 entries, made of 6,335 origin airports and 6,407 destination airports among 241 countries in the world.
Among these routes, a large amount have, historically, a maximal monthly volume of passengers (\texttt{maxP}) below 50 units in the period January 2010 - October 2019. We do not forecast or monitor these routes as their impact on the global aviation is very mild. Indeed, just as an example, the route HKG - TPE (Hong Kong - Taipei) has a \texttt{maxP} statistic equal to 229,316. 
After selecting the most frequnet routes (defined as: those routes with \texttt{maxP} $\geq$ 50) we end up with 222,557 routes, concerning 3,909 origin airports and 3,897 destination airports and involving 234 countries. This selected routes amount to, e.g.,  about 98.7\% of the volume of passengers in 2018 according to our data set.

\subsection{The Observed Scenario via On-Line Booking Data}\label{sec:kiwi}
 In order to have an estimate of the number of passengers lost due to observed route suppression, we use a different web-scraping strategy to look for all flights available corresponding to the routes considered in our data set. We used Kiwi\footnote{\texttt{Skypicker} are the API server of \texttt{Kiwi.com}}, which is a service used by travel agencies and individuals for on-line flight booking. Every week, or half week (since March), we inquire for the flights along a specific route (origin airport - destination airport) for the number of flights available for the next 7 days. We scan systematically all the 222,557 routes. It turns out, for example,  that around 1,000 unique connections from China have zero flights in the period 7-25 February 2020 and thus we take it as a proxy of the real number of routes temporary cancelled due to COVID-2019 outbreak. We will call later this the \texttt{Observed} scenario.

\subsection{Real-Time Flight Tracking Data}
Systems that track flights in real time can also be a valuable source of information about the operational status of aviation around the world. There are a number of private online platforms that aggregate this information and provide access to some it, either for free or on paid-subscription plans. These platforms receive aircraft positional reports from networks of terrestrial and space-based receivers as well as from air navigation service providers. Most modern large aircraft are visible in the tracking platforms, including both passenger and cargo planes. Although these systems do not give information about the number of passengers (or cargo) on the airplanes, they provide intelligence in almost real time about aircraft movements (e.g. departures, arrivals) as well as their tracks. 

In this study we are accessing data from OpenSky Network\footnote{\url{https://opensky-network.org}}. OpenSky provides open access to historical flight data but mostly for European and North America.
Compared to the on-line booking data, these type of data provide information for direct flights only but are timely and high-frequency. On the other side, booking scraping is based on the existence of some flight connection based on historical data so if a connection no longer exists, this is a clear sign of lock-down.

\section{The Forecasting Model}\label{sec:model}
To anticipate the release of more recent data from SABRE, we forecast the air volumes for the period November 2019 - December 2020 on the basis of historical trends between 2010 and October 2019. For the forecast we use a non-homogeneous Poisson process with a periodic intensity  function calibrated on historical data \citep{IacusYoshida2018} and extrapolated non-linearly for the future years, e.g., the time series of all January's from 2010 till 2019 is used to calibrate a regression model with quadratic terms and the value for 2020 is projected accordingly. The intensity function of the non-homogeneous Poisson process is therefore the combination over all periods. This approach is needed to  take into account that air passengers volumes show non-linear trends and cycles  which are specific to each route, so a model has been fitted for each route as well. Other approaches likes ARIMA or the alike can be used as well but we preferred to use pure counting processes rather than Gaussian processes.

With this approach we were able to forecast both air flows and (very roughly) overall ticketing revenues\footnote{More accurate impact assessment based on macro-economics analysis will be the object of a different forthcoming publication by JRC.} as if no flight/route were stopped. This will be called  the \texttt{Baseline} scenario\footnote{Remark that we have no information about the loading of the aircrafts, but in fact it is common sense that some flights were flying almost empty.}.
For the month of January 2020, as all flights from Wuhan were suppressed on January 23rd, therefore we rescaled the forecasted number of passengers accordingly along the routes originating from Wuhan anyway.

\section{Air Traffic scenarios}\label{sec:scenarios}
\begin{figure}[!htb]
    \centering\includegraphics[width=0.95\textwidth]{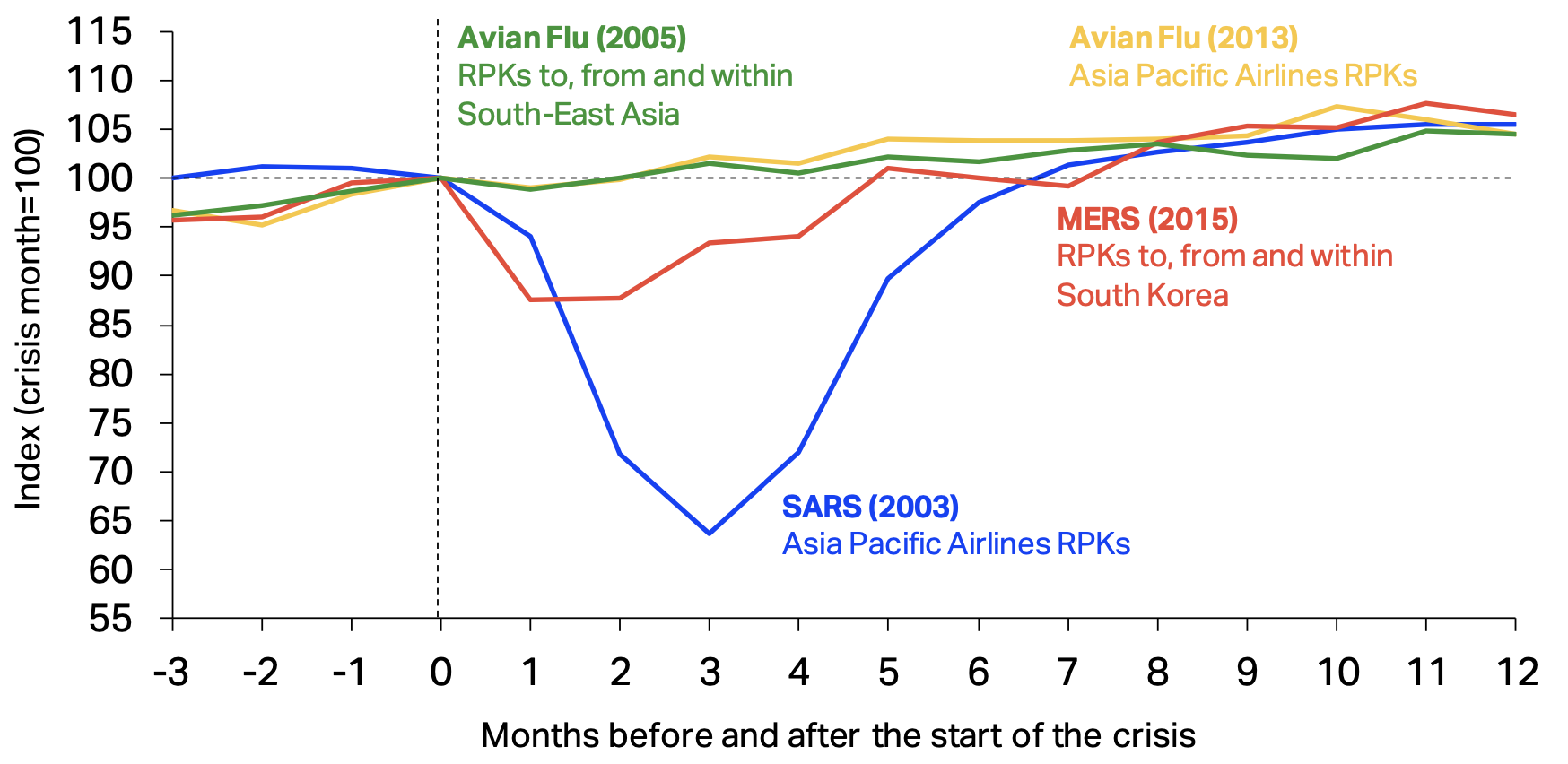}
    \caption{\textit{Impact of past outbreaks on aviation. Source IATA Economics.}}
    \label{figIATA}
\end{figure}

\begin{figure}[!htb]
    \centering\includegraphics[width=0.95\textwidth]{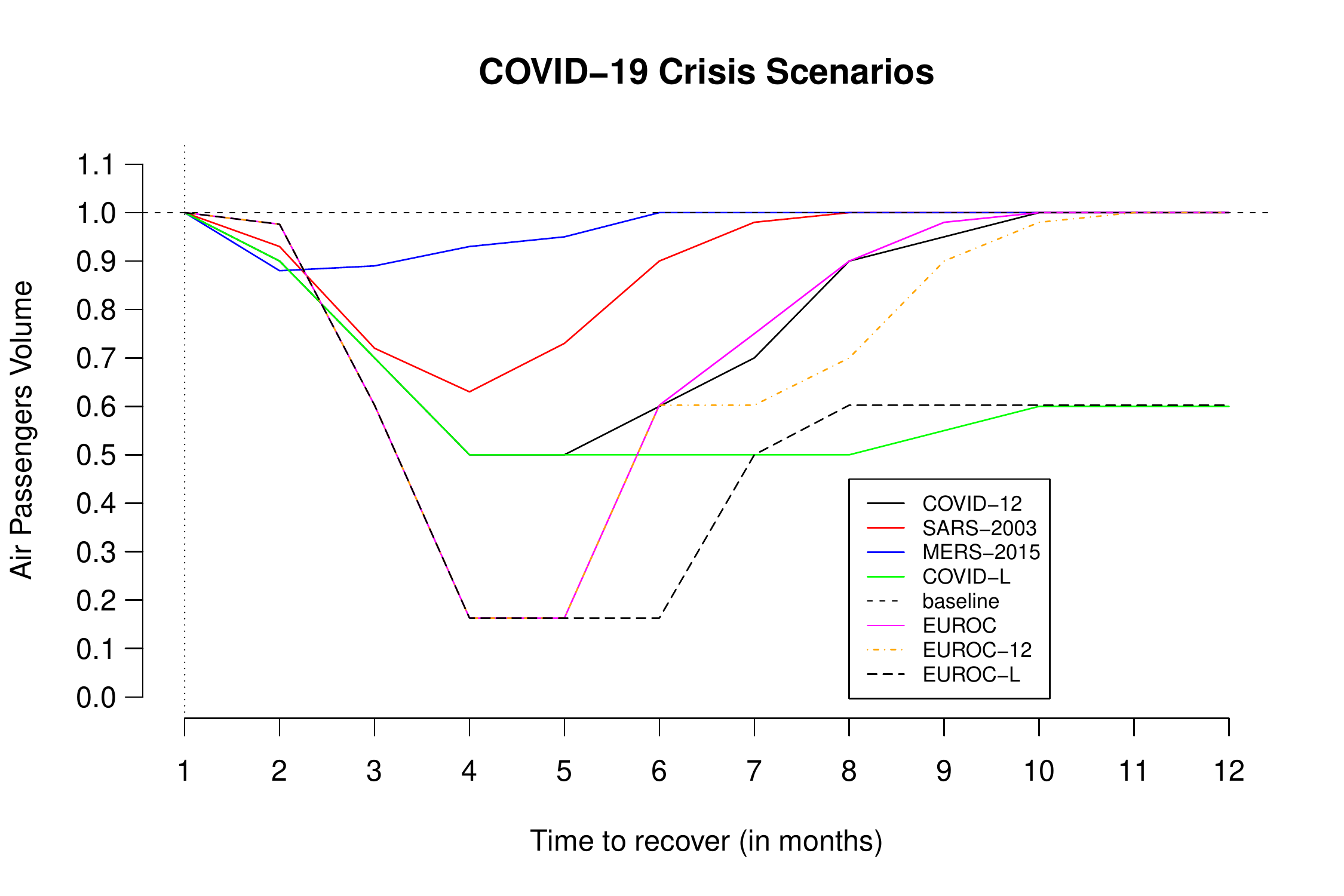}
    \caption{\textit{Hypothetical scenarios. On the x-axis, time in months (1 = January). On the vertical axis, proportion of volume of passengers (1 = pre-crisis volume). We ref to \texttt{Observed} and \texttt{EUROC} as the \textit{``most favorable''} scenarios, and to \texttt{COVID-L} and \texttt{EUROC-L} as the \textit{``worst case''} scenarios.}}
    \label{figScenarios}
\end{figure}

The forecasting model does not take into account the effects of coronavirus outbreak. We provide then several scenarios that try to mimic the past airline traffic disruption occurred in previous pandemic episodes and we called \texttt{Baseline} scenario the one based on simple forecast based on the model of Section~\ref{sec:model}. Figure~\ref{figIATA} shows the effect of past crisis \citep{IATApandemic} like the SARS-2003, MERS-2005 and others in terms of Revenue Passenger Kilometers\footnote{Revenue Passenger Kilometers (RPK) or Revenue Passenger Miles (RPM) is an airline industry metric that shows the number of kilometers traveled by paying passengers. It is calculated as the number of revenue passengers multiplied by the total distance traveled. Since it measures the actual demand for air transport, it is often referred to as airline ``traffic.''}
 (RPK) drop and the time to recover to the baseline status before the crises.  The scenarios we propose here only mimic the patterns but do not correspond exactly to the past scenarios for at least two reasons: \textit{i)} we consider air passenger volumes and not RPK and \textit{ii)} the above past scenario are quasi-local while we consider global scenarios.  In Figure~\ref{figIATA} we see the so-called  U-shaped pattern of the SARS-2003, which has been the most serious epidemic impacting traffic volumes in the recent period. At the height of the outbreak (May 2003), monthly RPKs of Asia-Pacific airlines were about 35\% lower than their pre-crisis levels. Overall in 2003, the loss of confidence and fears of global spread impacted both business and leisure travel to, from and within the region, resulting in Asia-Pacific airlines losing 8\% of annual RPKs and lasted about 9 months. The MERS Flu in 2015 produced a 12\% decline in monthly RPKs to, from and within South Korea in the first month of the outbreak. However, air travel volumes began to recover after two months and had returned to pre-outbreak levels within 6 months. On the other side, the two episodes of the avian flu in 2005 and 2013 did not affect the aviation market that much. 
As we know the COVID-19 is more global and impacting due to many restrictive measures imposed by countries and airlines, we will consider the \texttt{SARS-2003} and the \texttt{MERS-2015} patterns only for our scenario and we add two other scenarios: one, called \texttt{COVID-12}, which goes as low as 50\% and recovers in 12 months and the \texttt{COVID-L} (ell-shaped scenario) which is the same as \texttt{COVID} but never get back to the original volume and returns only at 60\%.
Figure~\ref{figScenarios} represents the different hypothesized scenarios. Another scenario, called the \texttt{Observed} scenario, is based on the analysis of the booking data presented in Section~\ref{sec:kiwi}. In this scenario, we discount the number of forecast passengers along a route if we observed no direct flights or no 1-stop flights, which means we assign zero passengers on that route. Up to March 2019 we are able to adjust this \texttt{Observed} scenario accordingly for all the routes monitored.
Finally we a set of additional three scenarios based on the Eurocontrol air traffic data \citet{eurocontrol} which provide global average air volumes up to date. The scenario are called \texttt{EUROC}, \texttt{EUROC-12} (recovering in 12 months) and \texttt{EUROC-L} (L-shaped version). 
There are two differences between the \texttt{EUROC*} scenarios and the \texttt{Observed} scenario:  \textit{i)} we do not project the \texttt{Observed} scenario after March 2020 and \textit{ii)} the observed scenario are calculated by route, while the \texttt{EUROC*} scenarios are global estimates.

Please, finally remark that we apply the scenarios at the same time, while in practice the flight bans are different from region to region.  Therefore, these scenarios should be thought as global average effects\footnote{In this respect, the \texttt{EUROC*}, being global averages, try to take into account these differences.}. The economic analysis in Section~\ref{sec:impact} will then be aggregated by quarter to take into account this shifting of scenarios across different regions.

To ease the description we refer to the \texttt{Observed} and \texttt{EUROC} as the \textit{``most favorable''}  scenarios, and to the \texttt{COVID-L} and \texttt{EUROC-L} as the \textit{``worst case''} scenarios.


\section{Global impact of travel ban on aviation, economy and society under the different scenarios}\label{sec:impact}
As mentioned in \citet{IATA2019}, \textit{``aviation's global stature as an economic engine is evident in the statistics. If the global aviation sector were a country, its total contribution (direct, indirect, induced and catalytic) of USD 2.7 trillion to the gross domestic product (GDP), and the 65.5 million jobs it supports, would be comparable to the United Kingdom's economic size and population''}.
In 2018, the official statistics report as much as 4.3 billion passengers, while in our SABRE data we calculated 3,773,950,899 (thus 87.7\% of the IATA official datum). In our selection of routes for which forecast and scenario calculation was possible, we observed 3,580,700,282 passengers, therefore 94.8\% of the whole SABRE data and 83.2\% of the official IATA datum. The total number of jobs supported by aviation is 65.5 million according to \citet{IATA2019}. According to recent estimates by the cross-industry Air Transport Action Group (ATAG), the total economic impact (direct, indirect, induced and tourism-connected) of the global aviation industry reached USD 2.7 trillion, about 3.6 per cent of the world’s gross domestic product (GDP) in 2016. On a global scale, the distribution of jobs aviation-related/induced and GDP impact is summarized in Table~\ref{tab:IATA2019}.
\begin{table}[!htb]   
    \centering
    \begin{tabular}{l|c|c|c|c}
       World &Jobs million & (\%)&  GDP billion & (\%)\\
       \hline
       Tourism catalytic &  35.7 & (56.0\%)& \$896.9 &  (33.3\%) \\
       Induced &  7.8 & (11.9\%) & \$454.0 & (16.9\%) \\
       Indirect &  10.8 & (16.5\%) & \$637.8 & (23.7\%)\\
       Aviation direct &  10.2 & (15.6\%) & \$704.4 & (26.2\%) \\
         \hline
        & 65.5 &  & \$2,693.1  &
    \end{tabular}
    \caption{\textit{Impact of the aviation sector on the economy.}}
    \label{tab:IATA2019}
\end{table}
Although the direct causal effect is beyond the scope of this short note, we can apply the different scenarios to roughly evaluate the potential socio-economic impact of the flight ban along 2020 due to the COVID-2019 crisis as well as the loss in terms of revenues as measured by airfares that we can observed in our SABRE data, also forecast according to the historical observed average fare per route trough simple seasonal regression model. In Table~\ref{tab:IATA2019}, the \textbf{direct impact} is about the  overall economic activity, jobs creation that directly serve passengers at airlines, airports and air navigation services providers. These include check-in, baggage handling, on-site retail,
cargo and catering facilities. But also include, jobs related to the manufacturing sector (those companies that produce aircraft, engines and other vital technologies). The \textbf{indirect impact} concerns the employment and economic activity generated by suppliers to the aviation industry: aviation fuel suppliers, etc. The \textbf{induced impact} is the spending of those directly or indirectly employed in the aviation sector supports additional jobs in other sectors such as retail outlets, companies producing consumer goods and a range of service industries (for example, banks, telecommunication providers and restaurants). Finally, \textbf{tourism catalytic} is related to air transport activities that affect multiple sectors of the economy, especially tourism and its value chain (hotels, restaurants, etc).

It is clear that flight ban affects all of the above in the short and long term. We leave very precise estimation of the above effects to further refinement of this research\footnote{Data on forecasting and scenarios are available to other experts in the field through the KCMD \href{https://bluehub.jrc.ec.europa.eu/migration/app/index.html}{Dynamic Data Hub} in section ``Mobility/COVID-19 Air traffic scenarios''.}. Here we will just apply our scenarios to the  figures in Table~\ref{tab:IATA2019} without any further ambition.
From the same \citet{IATA2019} report, we also see that around 90 per cent of business-to-consumer (B2C) e-commerce parcels are currently carried by air.

\begin{figure}[!htb]
    \centering\includegraphics[width=0.95\textwidth]{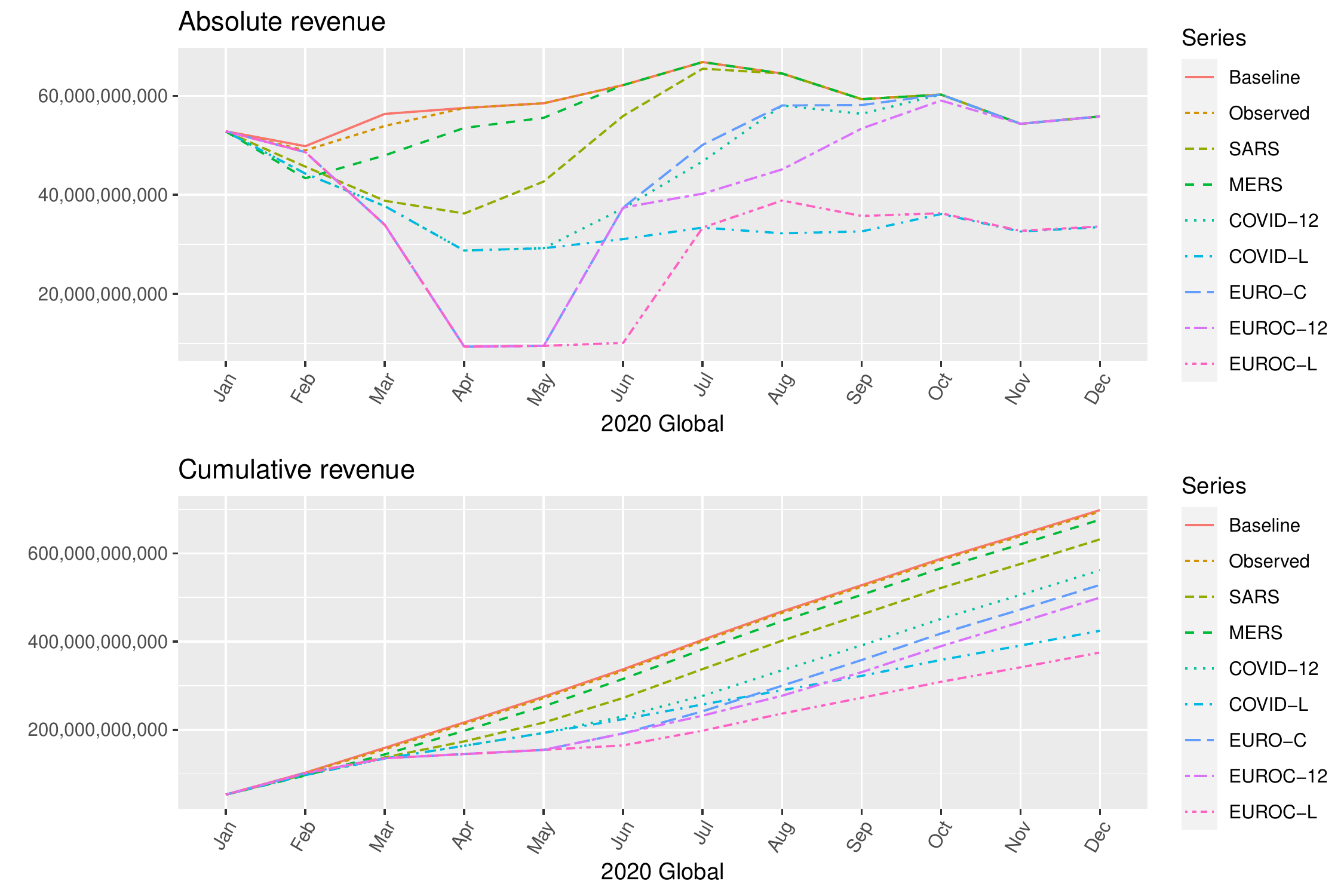}
    \caption{\textit{Impact on air fare loss under the different scenarios at global level in US\$.}}
    \label{figRevenueImpact}
\end{figure}

\begin{figure}[!htb]
    \centering\includegraphics[width=0.95\textwidth]{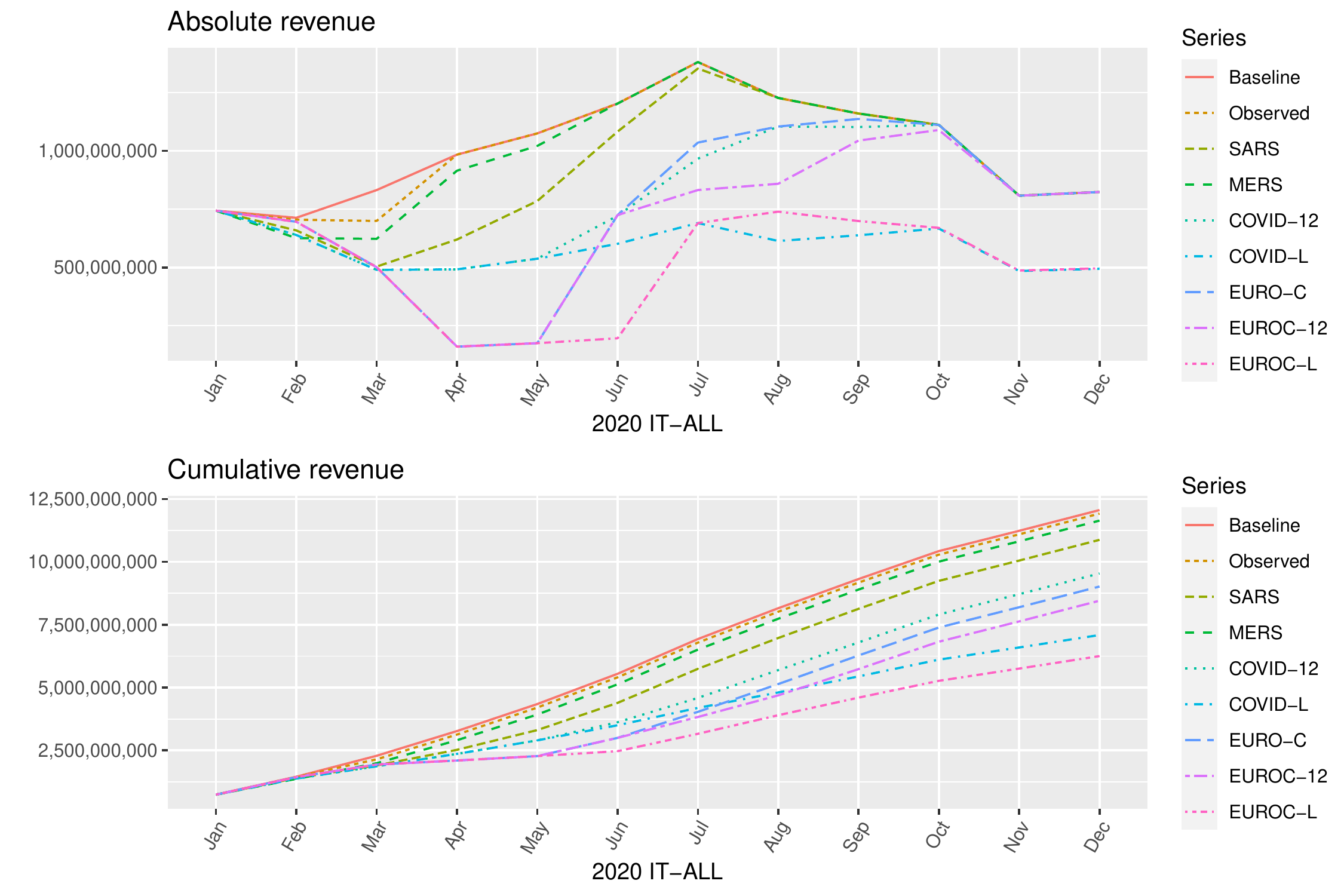}
    \caption{\textit{Impact on air fare loss under the different scenarios for all flights originating from Italy (in US\$). The \texttt{Observed} scenario is moving toward the other scenarios.}}
    \label{figRevenueImpactITIT}
\end{figure}

In Figure~\ref{figRevenueImpact} we show the direct impact on aviation due to loss of ticketing. On a global scale, we can notice only  a little difference between the \texttt{Baseline} scenario (nothing has happened) and the \texttt{Observed} scenario till the end of March. But if we look at specific situations, like e.g. Italy, we see that the observed number of flights decreases towards the hypothesized scenarios (see Figure~\ref{figRevenueImpactITIT}).
Table ~\ref{tab:globalLoss} estimates the loss by quarter compared to the baseline forecast. Values are in million of US\$. Assuming that a reduction of direct revenues / passengers may impact the entire aviation sector proportionally, we can roughly estimates the expected number of \textbf{job losses} and \textbf{impact on  GDP} at global scale. We assume that reduction has a 100\% impact, but clearly any other rescaling and more accurate factors will give more precise figures. The results are shown in Table~\ref{tab:globalLossGDP}. It turns out that for the first Quarter of 2020, the \texttt{Observed} travel ban impacts \textbf{0.02\% of the overall GDP} and \textbf{0.12\%} under the \texttt{EUROC} scenario, i.e., it has generated a loss of \textbf{\$3.3 billion} and \textbf{\$24 billion} respectively. Considering the whole year 2020,  in the worst scenario (\texttt{EUROC-L}) the global impact can go up to \textbf{1.67\%}, about \textbf{\$323 billion}. As tourism catalytic is 56\% of the global impact of aviation on the economy, one can rescale the results accordingly to estimate the impact on tourism-related impact.
\begin{table}[!htb]
\centering
\begin{tabular}{l|rrrrrrrr}
  \hline
 2020 & \texttt{Observed} & \texttt{SARS} & \texttt{MERS} & \texttt{COVID-12} & \texttt{COVID-L}& \texttt{EUROC} & \texttt{EUROC-12} & \texttt{EUROC-L} \\ 
  \hline
Q1 &  3,321.7 & 21,702.6 &   14,879 & 24,187.5 & 24,187.5 & 23,598.4 & 23,598.4 & 23,598.4 \\ 
  & (0.5\%) & (3.1\%) & (2.1\%) & (3.5\%) & (3.5\%) & (3.4\%) & (3.4\%) & (3.4\%) \\ 
  Q2 &          & 43,307.4 &  6,953.8 & 82,896.5 & 89,113.7 &  121,852 &  121,852 & 149,176.3 \\ 
  &   & (6.2\%) & (1.0\%) & (11.9\%) & (12.8\%) & (17.4\%) & (17.4\%) & (21.4\%) \\ 
  Q3 &         &  1,336.4 &          & 29,464.6 & 92,369.7 & 24,343.5 & 51,849.6 & 82,641.6 \\ 
   &   & (0.2\%) &   & (4.2\%) & (13.2\%) & (3.5\%) & (7.4\%) & (11.8\%) \\ 
  Q4 &          &          &          &           & 68,186.4 &         &  1,205.2 & 67,760.2 \\ 
   &   &   &   &   & (9.8\%) &   & (0.2\%) & (9.7\%) \\ 
     \hline
  Yearly &  3,321.7 & 66,346.4 & 21,832.8 & 136,548.6 & 273,857.2 & 169,793.9 & 198,505.2 & 323,176.6 \\  & (0.5\%) & (9.5\%) & (3.1\%) & (19.6\%) & (39.2\%) & (24.3\%) & (28.4\%) & (46.3\%) \\ 
\end{tabular}
\caption{\textit{Absolute and relative estimated loss for aviation in terms of ticketing under the different scenarios. The \texttt{Observed} scenario only take into accounts the current ban from end of January to end of March. In million of US\$.}}
\label{tab:globalLoss}
\end{table}
\begin{table}[!htb]
\centering
\begin{tabular}{l|rrrrrrrr}
  \hline
 2020 & \texttt{Observed} & \texttt{SARS} & \texttt{MERS} & \texttt{COVID-12} & \texttt{COVID-L} & \texttt{EUROC} & \texttt{EUROC-12} & \texttt{EUROC-L} \\ 
  \hline
  Q1 (jobs ml)  & \textbf{0.31} & 2.04 & 1.40 & 2.27 & 2.27 & \textbf{2.21} & 2.21 & 2.21 \\ 
  Q2 (jobs ml) &  & 4.06 & 0.65 & 7.77 & 8.36 & 11.43 & 11.43 & 13.99 \\ 
  Q3 (jobs ml) &  & 0.13 & 0.00 & 2.76 & 8.66 & 2.28 & 4.86 & 7.75 \\ 
  Q4 (jobs ml) &  & 0.00 & 0.00 & 0.00 & 6.39 & 0.00 & 0.11 & 6.35 \\ 
  Yearly (jobs ml)  &  & 6.22 & 2.05 & 12.81 & \textbf{25.68} & 15.92 & 18.62 & \textbf{30.31} \\ 
  \hline
  \hline
Q1 (GDP bl)&    12.81 &    83.69 &    57.37 &    93.27 &    93.27 &       91 &       91 &       91 \\ 
  Q2 (GDP bl)&          &   166.99 &    26.81 &   319.65 &   343.62 &   469.86 &   469.86 &   575.22 \\ 
  Q3 (GDP bl)&          &     5.15 &          &   113.62 &   356.18 &    93.87 &   199.93 &   318.67 \\ 
  Q4 (GDP bl)&         &         &          &          &   262.93 &          &     4.65 &   261.28 \\ 
  Yearly &     &   255.83 &    84.19 &   526.53 & 1,055.99 &   654.73 &   765.44 & 1,246.17 \\ 
   \hline
  \hline
  Q1 (World GDP \%)& \textbf{0.02} & 0.11 & 0.08 & 0.12 & 0.12 & \textbf{0.12} & 0.12 & 0.12 \\ 
  Q2 (World GDP \%)&   & 0.22 & 0.04 & 0.43 & 0.46 & 0.63 & 0.63 & 0.77 \\ 
  Q3 (World GDP \%)&   & 0.01 &   & 0.15 & 0.48 & 0.13 & 0.27 & 0.43 \\ 
  Q4 (World GDP \%)&   &   &   &   & 0.35 &   & 0.01 & 0.35 \\ 
  Yearly (World GDP \%)&   & 0.34 & 0.11 & 0.70 & \textbf{1.41} & 0.88 & 1.02 & \textbf{1.67} \\ 
\end{tabular}
\caption{\textit{Estimated job and GDP losses under the different scenarios assuming a proportional impact. Jobs in millions and GDP in billion of US\$ and percentage of global GDP loss, respectively.}}
\label{tab:globalLossGDP}
\end{table}
In terms of impact on the society, we can see from Table~\ref{tab:globalLossGDP} that the number of \textbf{job losses} may vary by \textbf{310,000 units} for the first quarter of 2020 under the \texttt{Observed} scenario and \textbf{2.2 million} under the \texttt{EUROC} scenario, and up to \textbf{30.3 million} worldwide under the worst scenario \texttt{EUROC-L}.

\section{EU27 regional impact}\label{sec:europe}
\begin{figure}[!htb]
    \centering\includegraphics[width=0.95\textwidth]{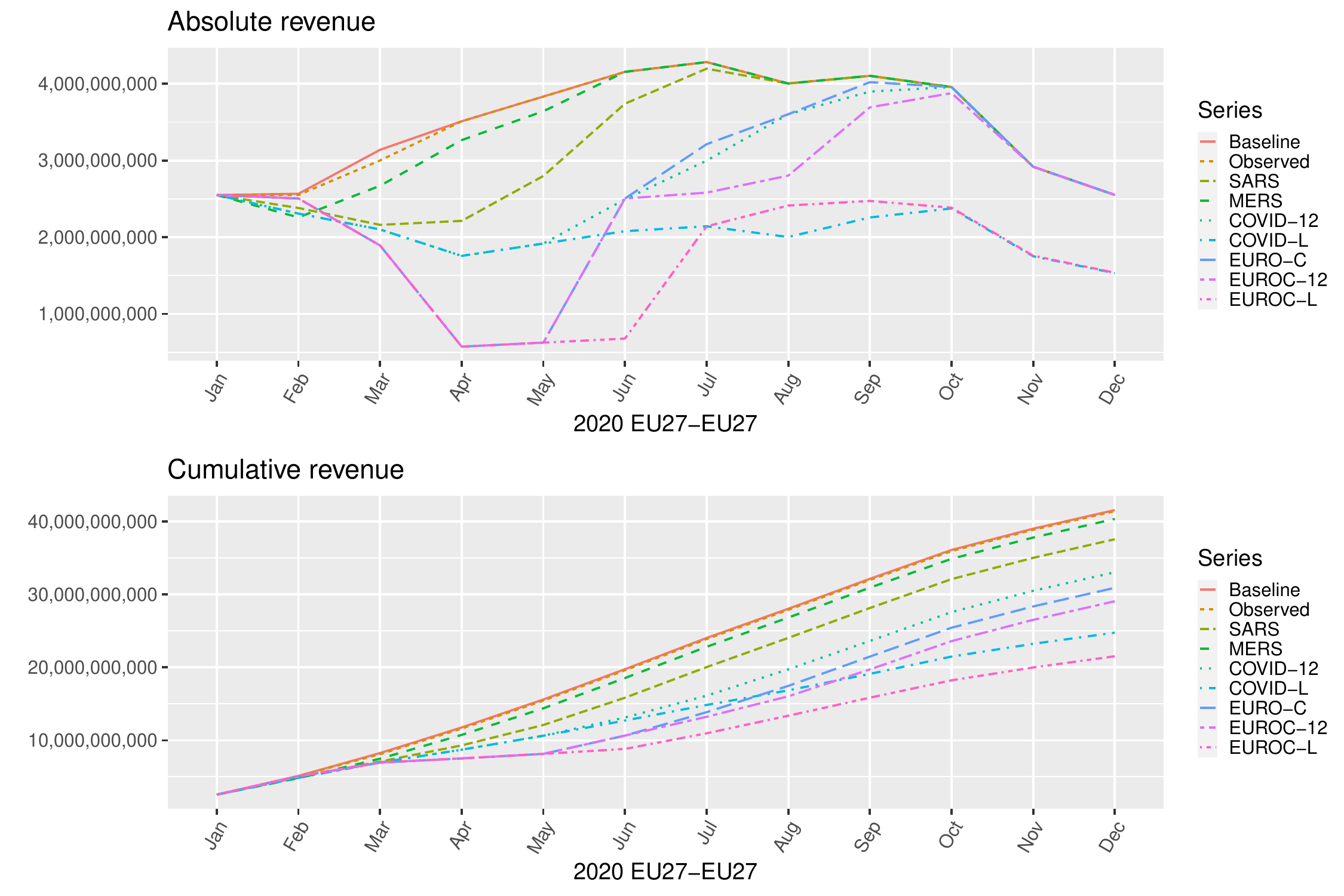}
    \caption{\textit{Impact on air fare loss under the different scenarios for flights from all EU27 to all EU27 countries (in US\$).}}
    \label{figRevenueImpactEUEU}
\end{figure}

\begin{figure}[!htb]
    \centering\includegraphics[width=0.95\textwidth]{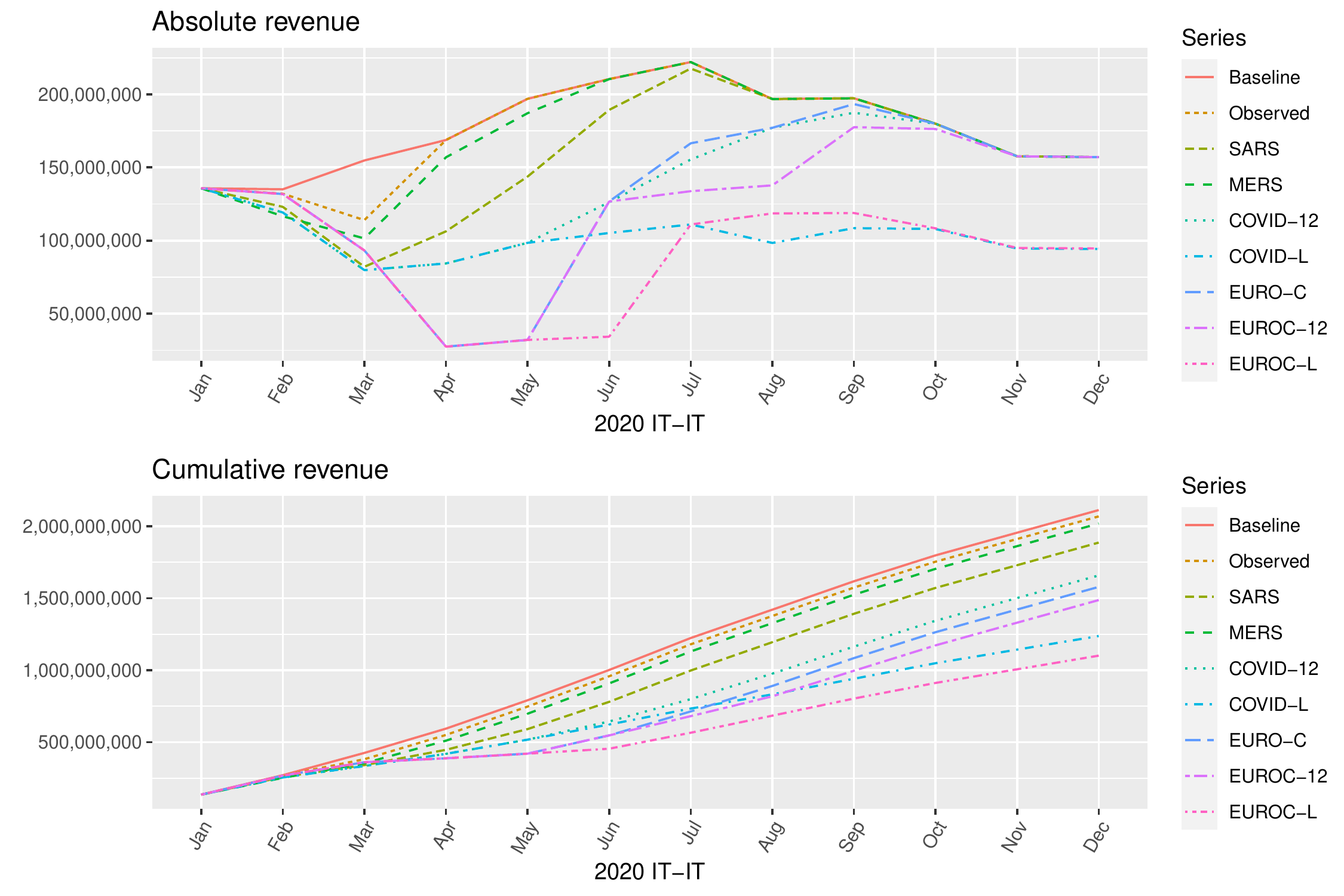}
    \caption{\textit{Impact on air fare loss under the different scenarios for all domestic flights in Italy (in US\$).}}
    \label{figRevenueImpactIT}
\end{figure}

In \citet{ATAG2018} the impact of aviation is also analyzed by regions: Africa, Asia and Pacific, Europe, Latin America and Caribbean, Middle East and North America. Table~\ref{tab:IATAEU} reports the rescaled\footnote{As of February 2020, UK is no longer part of the EU hence we rescaled the IATA aviation statistics proportionally to the  amount of EU27 over the former EU28 total GDP, which is about 85\%.} data for the EU27 for which the impact of aviation on GDP is about 4.1\% and the total number of jobs amounts to 3.3\%, according to the latest statistics.  Considering only flights from EU27 to EU27 countries (see Figure~\ref{figRevenueImpactEUEU}),  it turns out that these very rough figures are comparable to those reported in \citet{JRCB5}: in the first Quarter of 2020, the \texttt{Observed} travel ban impacts \textbf{0.02\%} of the EU27 GDP and \textbf{0.13\%} under the \texttt{EUROC} scenario, while in the \textbf{worst scenario} (\texttt{EUROC-L}) the global impact can go up to \textbf{1.98\%} on a yearly base. In terms of job losses, in the first Quarter of 2020, the \texttt{Observed} and \texttt{EUROC} scenario report \textbf{40,000} and \textbf{330,000} job losses respectively. For the whole 2020, under the worst scenario (\texttt{EUROC-L}) this number rises to \textbf{5 million} of job losses.
Looking at Figure~\ref{figRevenueImpactEUEU} and Figure~\ref{figRevenueImpactIT} it emerges that the \texttt{Observed} scenario has different shape locally and globally at EU27 level.
\begin{table}[!htb]   
    \centering
    \begin{tabular}{l|c|c|}
       EU27 &Jobs &    GDP   \\
        &million &     billion \\
       \hline
       Tourism catalytic &  4.3  & \$249.6   \\
       Induced &  1.3   & \$94.7   \\
       Indirect &  2.6   & \$191.7  \\
       Aviation direct &  2.2 &   \$163.7   \\
         \hline
        & 10.4  &  \$669.6   
    \end{tabular}
    \caption{\textit{Impact of aviation in EU27.}}
    \label{tab:IATAEU}
\end{table}
\begin{table}[!htb]
\centering
\begin{tabular}{l|cccccccc}
  \hline
 2020 & \texttt{Observed} & \texttt{SARS} & \texttt{MERS} & \texttt{COVID-12} & \texttt{COVID-L} & \texttt{EUROC} & \texttt{EUROC-12} & \texttt{EUROC-L} \\ 
  \hline
  Q1 &    154.3 &  1,164.7 &    776.9 &  1,297.6 &  1,297.6 &  1,309.6 &  1,309.6 &  1,309.6 \\ 
  & (0.4\%) & (2.8\%) & (1.9\%) & (3.1\%) & (3.1\%) & (3.1\%) & (3.1\%) & (3.1\%) \\ 
    Q2 &          &  2,750.1 &    437.5 &  5,335.3 &    5,751 &  7,800.4 &  7,800.4 &  9,627.1 \\ 
  &  & (6.6\%) & (1.1\%) & (12.8\%) & (13.8\%) & (18.8\%) & (18.8\%) & (23.2\%) \\ 
 Q3 &          &     85.7 &         &  1,890.7 &  5,990.6 &  1,553.4 &  3,314.4 &  5,364.7 \\ 
   &   & (0.2\%) &  & (4.5\%) & (14.4\%) & (3.7\%) & (8\%) & (12.9\%) \\ 
    Q4 &          &         &         &         &  3,770.9 &         &     79.2 &  3,747.3 \\ 
   Q4 &   &   &   &   & (9.1\%) &   & (0.2\%) & (9\%) \\ 
\hline
 Yearly &    154.3 &  4,000.4 &  1,214.4 &  8,523.6 & 16,810.1 & 10,663.4 & 12,503.5 & 20,048.8 \\ 
   & (0.4\%) & (9.6\%) & (2.9\%) & (20.5\%) & (40.4\%) & (25.6\%) & (30.1\%) & (48.2\%) \\ 
 \end{tabular}
\caption{\textit{Estimated loss in ticketing under the different scenarios for the EU27. The \texttt{Observed} scenario only take into accounts the current ban from end of January to end of March. In million of US\$.}}
\label{tab:EULoss}
\end{table}
\begin{table}[!htb]
\centering
\begin{tabular}{l|rrrrrrrr}
  \hline
 2020 & \texttt{Observed} & \texttt{SARS} & \texttt{MERS} & \texttt{COVID-12} & \texttt{COVID-L} & \texttt{EUROC} & \texttt{EUROC-12} & \texttt{EUROC-L} \\ 
  \hline
  Q1 (jobs ml)& \textbf{0.04} & 0.29 & 0.19 & 0.32 & 0.32 & \textbf{0.33} & 0.33 & 0.33 \\ 
  Q2 (jobs ml)&   & 0.69 & 0.11 & 1.33 & 1.43 & 1.95 & 1.95 & 2.40 \\ 
  Q3 (jobs ml)&   & 0.02 &  & 0.47 & 1.49 & 0.39 & 0.83 & 1.34 \\ 
  Q4 (jobs ml)&     &   &   &  & 0.94 &  & 0.02 & 0.93 \\ 
  Yearly (jobs ml) && 1.00 & 0.30 & 2.13 & \textbf{4.19} & 2.66 & 3.12 & \textbf{5.00}\\
  \hline
  \hline
  Q1 (GDP bl) &          2.60 &     19.6 &    13.07 &    21.84 &    21.84 &    22.04 &    22.04 &    22.04 \\ 
  Q2 (GDP bl) &           &    46.28 &     7.36 &    89.78 &    96.78 &   131.27 &   131.27 &   162.01 \\ 
  Q3 (GDP bl) &           &     1.44 &         &    31.82 &   100.81 &    26.14 &    55.77 &    90.28 \\ 
  Q4 (GDP bl) &           &         &          &          &    63.46 &          &     1.33 &    63.06 \\ 
  Yearly (GDP bl) &   2.60 &    67.32 &    20.44 &   143.44 &   282.88 &   179.44 &   210.41 &   337.38 \\  
  \hline
  \hline
  Q1 (EU27 GDP \%)& 
   \textbf{0.02} & 0.11 & 0.08 & 0.13 & 0.13 & \textbf{0.13} & 0.13 & 0.13 \\ 
  Q2 (EU27 GDP \%)&     & 0.27 & 0.04 & 0.53 & 0.57 & 0.77 & 0.77 & 0.95 \\ 
  Q3 (EU27 GDP \%)&     & 0.01 &   & 0.19 & 0.59 & 0.15 & 0.33 & 0.53 \\ 
  Q4 (EU27 GDP \%)&   &   &   &   & 0.37 &   & 0.01 & 0.37 \\ 
  Yearly (EU27 GDP \%)&   & 0.39 & 0.12 & 0.84 & \textbf{1.66} & 1.05 & 1.23 & \textbf{1.98} \\
 \end{tabular}
\caption{EU27: Estimated job and GDP losses under the different scenarios assuming a proportional impact. Jobs in million and GDP in billion of US\$ and percentage of global GDP loss, respectively.}
\label{tab:EULossGDP}
\end{table}

\section{Analysis of Real-Time Traffic on Selected Airports}\label{sec:tracking}

This section presents an assessment of the first changes in flight activity around the world as a result of the COVID-19 pandemic. The analysis uses airplane movements from the online flight tracking platform OpenSky Network \citep{OpenSky} to estimate alterations in activity that are possibly linked to a reduction in the demand for international travel and to mobility restriction measures imposed by authorities. To this end, we collected aircraft departures (i.e. take-offs) from January to March 2020 for 141 airports. The criteria for selecting these 141 airports was a balance between air passenger traffic, geographical coverage and data availability. OpenSky Network was selected since it provides open access to historical flight data. However, the limitation of this source is that it does not provide global coverage and it cover mostly for Europe and North America. For example, flight data for major global airports such as the Beijing Capital International Airport in China and the Soekarno–Hatta International Airport in Indonesia are not available. 

In this light, during recent weeks traffic has reduced significantly in all European countries, while it has remained fairly stable in Far East Asia.

\begin{figure}[!htb] 
    \centering
    \includegraphics[width=\textwidth]{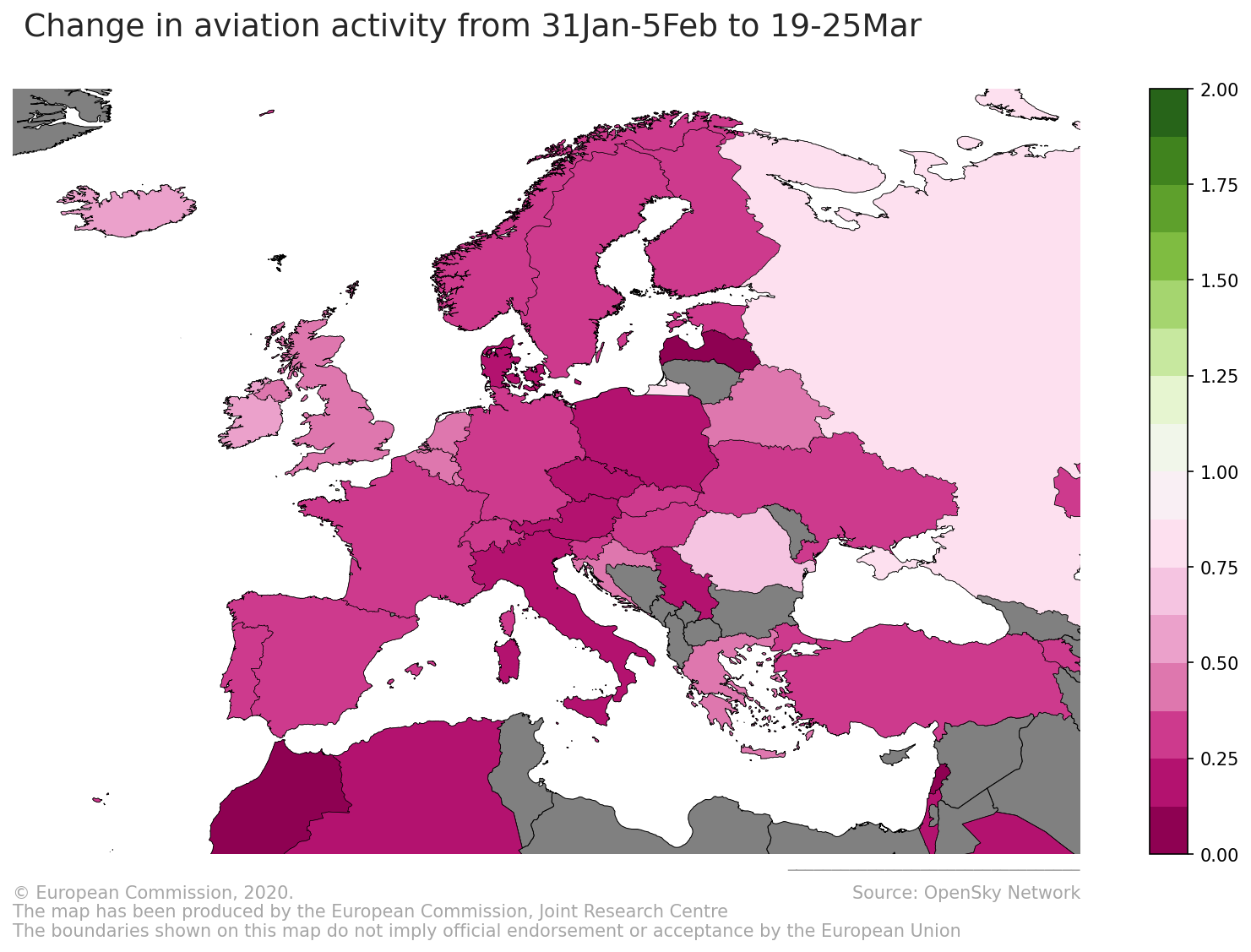}
    \caption{Number of flight departures during 19 to 25 March 2020, compared to the number of departures during 30 January to 5 February 2020. Green colour indicates an increase in flight numbers, pink a decrease. Source: OpenSky Network}
    \label{fig:map_europe}
\end{figure}

The map in Figure~\ref{fig:map_europe} compares the number of flight departures from a given country during  19 to 25 March 2020, with the number of departures six weeks earlier (30 January to 5 February). In Europe, all countries experienced a sizable decline in activity. The number of flights in a country is the result of aggregating the flights from all the airports we monitor in that country. It is clear that we only capture some of the aviation activity in the country, but by including in the analysis the busiest airports, we believe the estimates we provide are representative of the overall activity in the countries. OpenSky tracks both passenger and cargo airplanes, and we make no attempt to separate these two types of traffic. So the estimates presented in this and following sections refer to airplane movements - specifically departures -, not to passenger airplane movements, or to number of passengers or to number of available airline seats.In the map, the lower the value for a country, the higher the reduction of flights in that country. Values above 1 (green colour on the map) indicate an increase in the number of flights.   

\begin{figure}[!htb] 
    \centering
    \includegraphics[width=\textwidth]{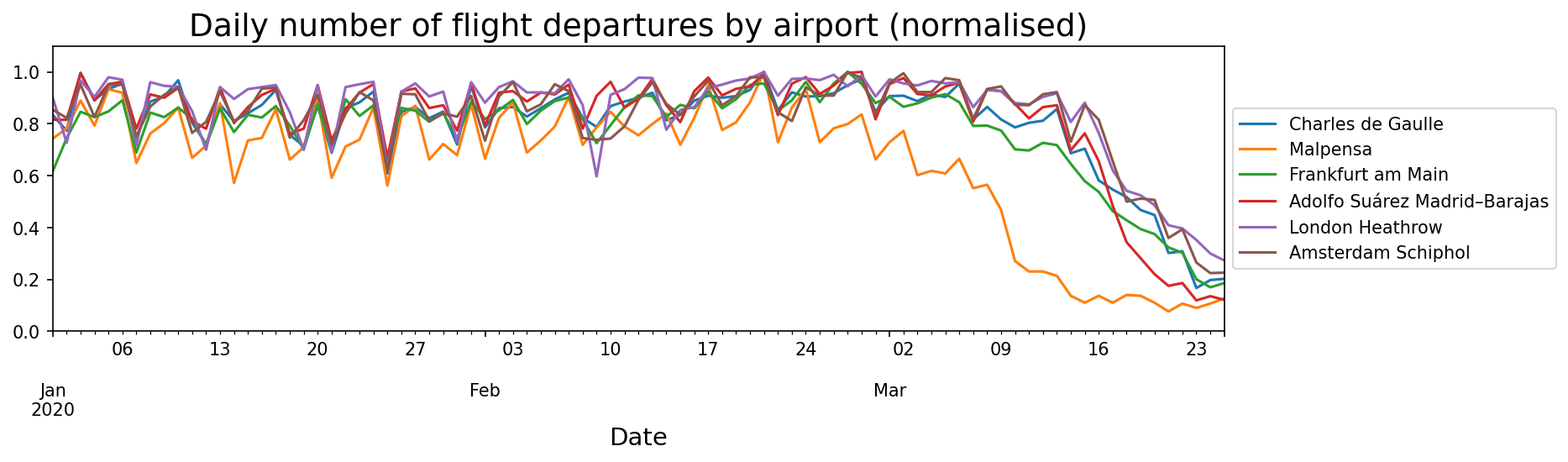}
    \caption{Daily number of flight departures at some European airports (normalised data). Source: OpenSky Network}
    \label{fig:airports_Europe}
\end{figure}

\begin{figure}[!htb] 
    \centering
    \includegraphics[width=\textwidth]{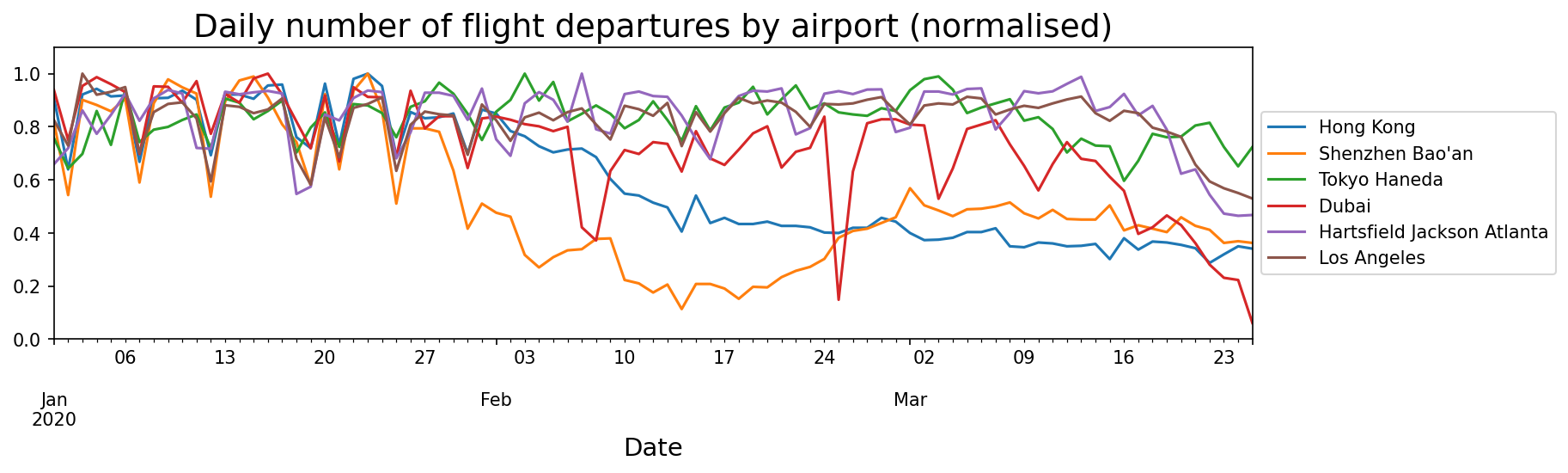}
    \caption{Daily number of flight departures at some airports in the rest of the world (normalised data). Source: OpenSky Network}
    \label{fig:airports_RestofW}
\end{figure}

Trends analysis of the number of departures can give insights as to the impact that a reduced demand for travelling or measures like lockdowns are having on aviation activity. Figure~\ref{fig:airports_Europe} shows the trend in daily flight departures at some major European airports during the period 1 January to 25 March 2020. The numbers have been normalised to the maximum number of departures in that time period. The earliest reduction in flights is seen at Malpensa airport (Milan); the rest of the airports have followed the same pattern with a delay of a few days. 
Figure~\ref{fig:airports_RestofW} shows the flight trends for some airports in the rest of the world. The airport of Shenzhen has actually recovered in February and March some of the activity lost earlier in the year. The drops in the other airports are less severe than in Europe, with the exception of Dubai airport, which has experienced a sharp decline since 15 March

\begin{figure}[!htb]
    \centering
    \includegraphics[width=\textwidth]{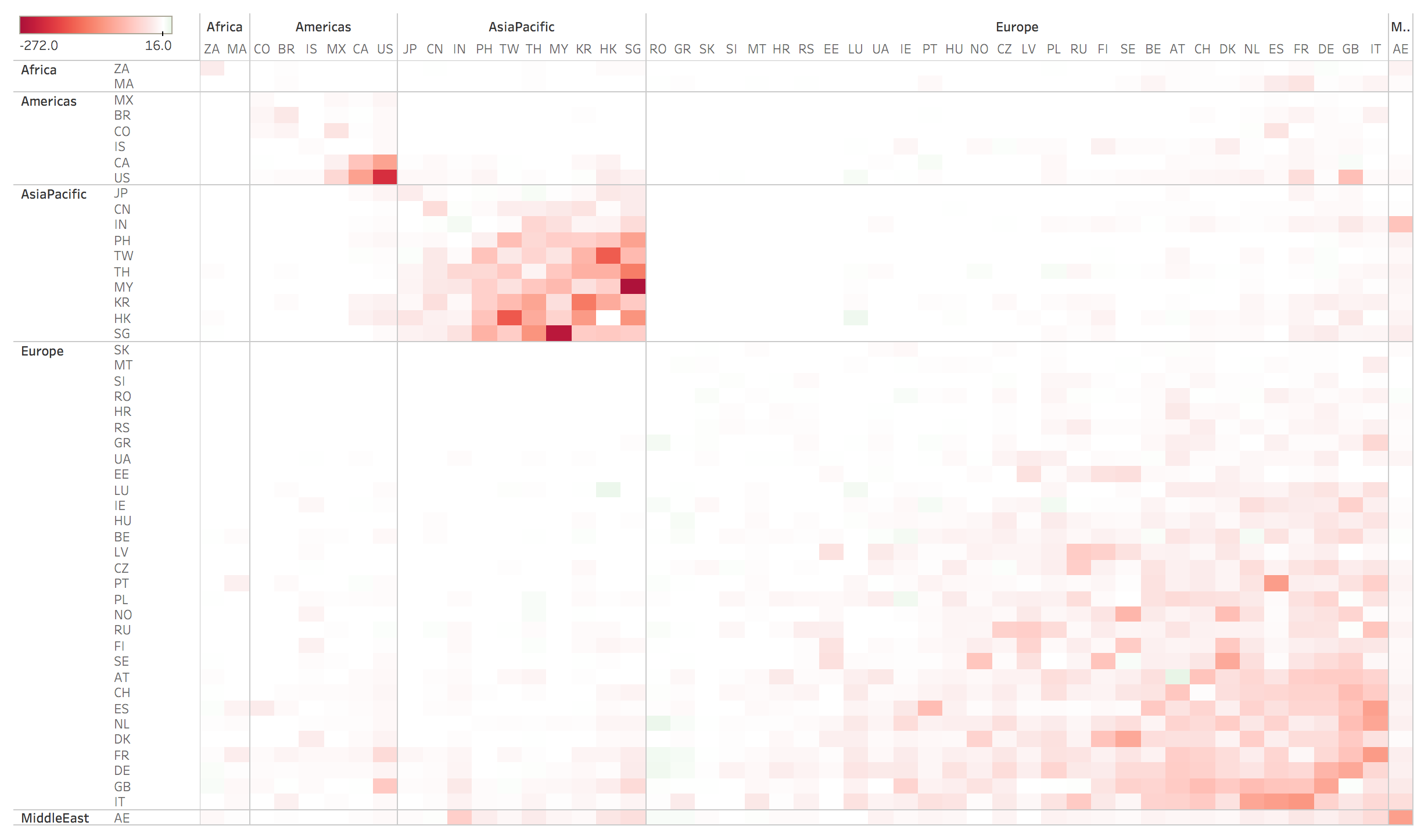}
    \caption{Difference between the number of flight departures during 19 to 25 March 2020, compared to the number of departures during 30 January to 5 February 2020 by country of origin (vertical axis) and destination (horizontal axis). Green colour indicates an increase in flight numbers, red a decrease. Data for 49 selected countries. Source: OpenSky Network}
    \label{fig:flight_bilateral}
\end{figure}

The heat-map in Figures~\ref{fig:flight_bilateral} compares the number of flight departures from a given country to a given destination country during 19 to 25 March 2020, compared to the number of departures during 30 January to 5 February 2020. For this heat-map, we selected to visualize 49 countries using flight data from 54 out of the total 141 airports for which we collected from the OpenSky Network. During the compared time periods, the number of international flights was decreased significantly in all the continents.

\begin{figure}[!htb]
    \centering
    \includegraphics[width=\textwidth]{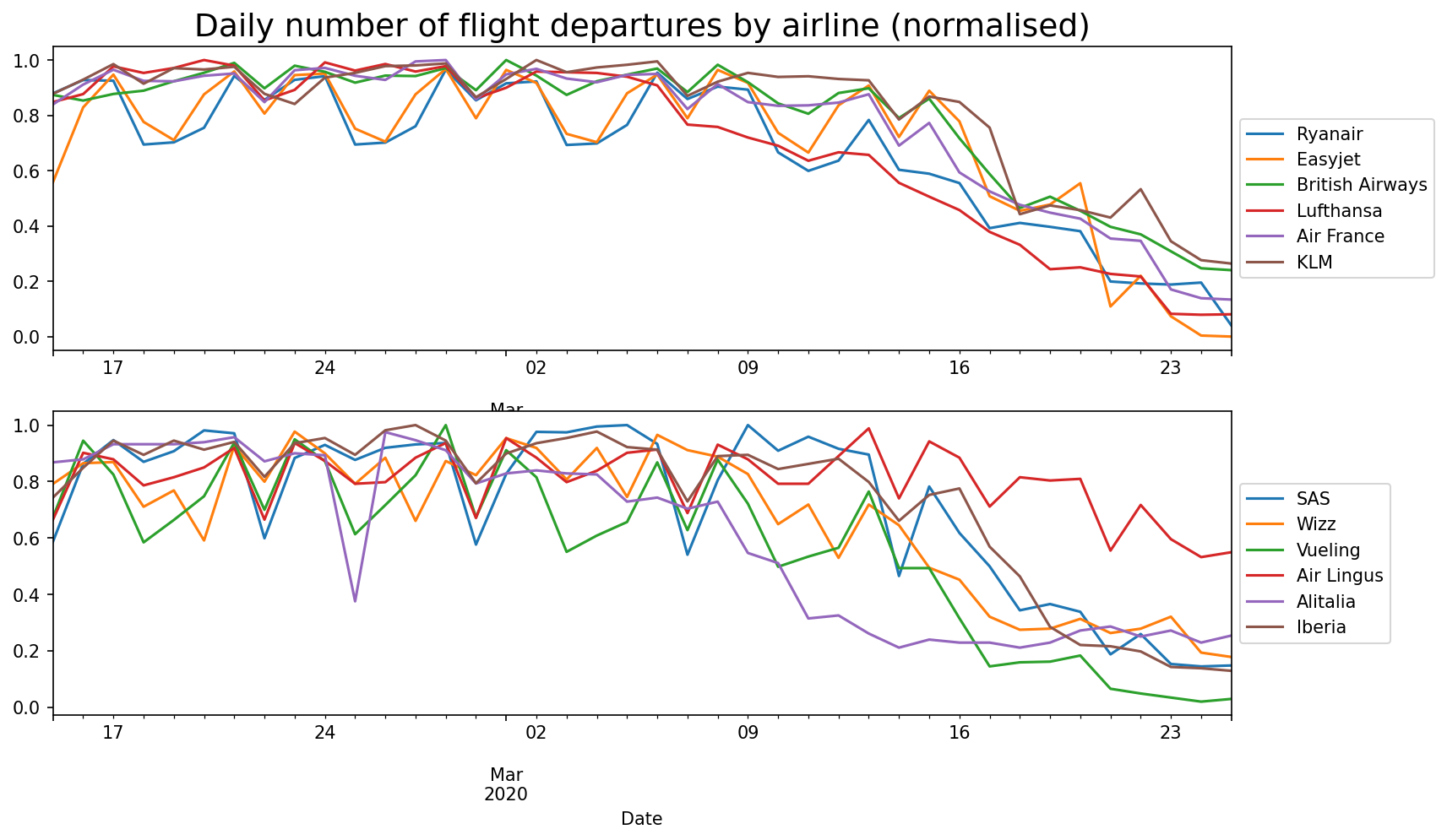}
    \caption{Number of flights for a subset of European airlines (top 12 in our data) from 15 Feb till 25 March, 2020. The numbers have been normalised to the maximum number of departures in that time period. Source: OpenSky Network}
    \label{fig:Airlines}
\end{figure}

Finally, for the most frequent airlines in our flight tracking data we were also able to estimate the week-by-week flight decrease per airlines.
It turns out that the major European airlines have dropped, if not shut down completely, the operations with obvious implications which are not addressed in this report. Figure~\ref{fig:Airlines} is self-explanatory.

\section{Caveats of the Analysis or What Can Go Wrong?}
While the modeling exercise used to forecast the air traffic is quite robust and has been back tested extensively, the projection and the choice of the different scenarios as well as the rough evaluation of the direct socio-economic impact has limits in its simplified approach. Nevertheless, the study seems to portrait not-unrealistic figures.
Among the caveats we list the following:
\begin{itemize}
   \item the duration of the Covid-19 pandemic is not known. \citet{Kissler2020} suggested that \textit{``The overall duration of the SARS-CoV-2 epidemic could last into 2022''}. In our analysis we assume that, in most scenarios, by December 2020 the aviation recovers its previous business levels;
\item measures such as national and regional lock-downs have a major impact on the aviation and the temporal length of these measures  cannot be foreseen and may vary across different regions and countries. Countries have different lock-down strategies from extreme lock-downs as in China/Wuhan, mild lock-downs in Europe and (early) light lock-downs in the UK, USA. Lock-downs aims at postponing the epidemic wave (``flattening the curve'') to reduce pressure on the health system. Different lock-down strategies impact the duration of pandemic and thus the impact on the aviation. Our scenarios should be intended as the result of an average of all these measures on a global scale;
 \item in our scenarios we assume that no second wave of Sars-CoV-2 epidemic will happen building on the past experience from Sars-2003 and Mers-2013. However, this is a new virus and could have a different seasonal behaviour;
 \item the start and the end of the impact of Covid-19 pandemic on aviation differs and it will differ (given the different strategies in place) across different countries. Once again, our scenarios should be considered as average effects.
 \item the multipliers used to evaluate the impact of aviation on GDP and jobs are quite approximated as we do not take into account any macro-economic (supply/demand) model. A sector by sector analysis is required to have precise figures. Nevertheless, the emerging figures are in line with other simulations \citep[e.g.][]{JRCB5}.
\end{itemize}

\section{Conclusions}
This work prepares the forecast of air traffic passengers and ticket fares for more than 222,557 routes around the world, concerning 3,909 origin airports and 3,897 destination airports and involving 234 countries, consisting in about 98.7\% of the volume of passengers in 2018 according to our SABRE data set. 

These forecast are then discounted, through different scenarios, for the flight ban occurring since last January 2020 around the world. The scenarios are based on both observed routes and flights cancellation using a mix of flight tracking data and on-line booking data, as well as hypotheses based on previous pandemic experience that affected aviation.

Under these projections, we then try to calculate the economic impact measured in terms of loss of GDP due to the aviation sector as well as the social impact due to job losses related to aviation and correlated sectors (tourism, catering, etc). Although elementary, these scenarios provide interesting figures in line with some other studies.

The forecasting and scenarios data for the year 2020,  aggregated by countries of origin and destination, are available through the KCMD Dynamic Data Hub\footnote{\url{https://bluehub.jrc.ec.europa.eu/migration/app/index.html\#?state=5e9ef39c60ab3f019a77a669}} for further analysis by the research community and on-line visualization. The complete forecast and scenario data set of airport-to-airport forecast are available on requests to the authors.

\bibliography{refbib}

\bibliographystyle{chicago}

\end{document}